\documentclass[manuscript]{aastex}

\slugcomment{Submitted to Astrophysical Journal Letters}

\shorttitle{Titan-Hyperion Resonance}
\shortauthors{\' Cuk, Dones \& Nesvorn\' y }

\begin{document}

\title{Titan-Hyperion Resonance and the Tidal Q of Saturn}

\author{Matija \' Cuk}
\affil{Carl Sagan Center, SETI Institute, \\
 189 North Bernardo Avenue, Mountain View, CA 94043\\
\email{mcuk@seti.org}}

\author{Luke Dones} \and \author{David Nesvorn\' y}
\affil{Southwest Research Institute, \\
Boulder, CO 80302}

\begin{abstract}
Lainey et al. (2012), by re-analyzing long-baseline astrometry of Saturn's moons, have found that the moons' tidal evolution is much faster than previously thought, implying an order of magnitude stronger tidal dissipation within Saturn. This result is controversial and implies recent formation of at least some of the mid-sized icy moons of Saturn. Here we show that this more intensive tidal dissipation is in full agreement with the evolved state of the Titan-Hyperion resonance. This resonance was previously thought to be non-tidal in origin, as the amount of tidal evolution required for its assembly is beyond what is possible in models that assume that all the major moons are primordial. We find that the survival of the Titan-Hyperion resonance is in agreement with a past Titan-Iapetus 5:1 resonance, but not with unbroken tidal evolution of Rhea from the rings to its current distance.
\end{abstract}


\section{Introduction}

\defcitealias{lai12}{L12}

Saturn's regular satellite system is widely thought to have formed from a flat disk coplanar with Saturn's rings, probably soon after formation of the planet \citep[but see][]{asp13, ham13}. After formation, the moons evolved outward by varying amounts due to tidal dissipation within Saturn, forming mean-motion resonances in the process \citep{md99}. Since \citet{gol65}, the assumption that the present icy satellites (from Mimas to Rhea) are primordial has set the upper limit on tidal dissipation of Saturn equivalent to $Q > 18,000$ \citep{pea80, mey07}\footnote{The upper limit on dissipation is equivalent to a lower limit on the tidal quality factor $Q$.}. If the tidal $Q$ of Saturn were lower, Tethys would have evolved past its present orbit over 4.5 Gyr. However, the ``classical" picture of the Saturnian system leaves many unanswered questions, three of which are:

1) The 4:3 mean motion resonance between Titan and Hyperion. The amount of outward tidal evolution by Titan required to bring the resonance to the present state is about an order of magnitude too large compared to what is expected if $Q > 18,000$ \citep{gre73, col74, lee00}. The classical solution is that the resonance must be primordial \citep[i.e., it was assembled during the moons' formation;][]{gol65, col74}, but it is unclear if dissipation in the protosatellite nebula could have evolved the resonance without damping the eccentricity of Hyperion.

2) The eccentricity of Titan of $e \simeq 0.03$ is too high, as tidal dissipation within Titan should damp its eccentricity \citep{sag82, soh95}. 


3) Iapetus exhibits an eight degree inclination to its Laplace plane, and the origin of this tilt is unknown. 


Recently, \citet{lai12} (hereafter L12) published an analysis of historical astrometry for Saturn's mid-sized icy satellites. They find evidence for strong tidal dissipation within Saturn, equivalent to $Q/k_2 \simeq 4000 -5000$.\footnote{$k_2$ is the tidal Love number of Saturn, thought to be about $0.34$ \citep{gav77}.} \citetalias{lai12} note that this fast tidal evolution could explain the observed heat flux of Enceladus in a straightforward manner \citep{how11}. A fast-evolving Enceladus would be advancing faster thorugh the resonance with Dione, and stronger satellite tides would be required to maintain an equilibrium, producing the observed tidal heating in the process \citepalias{lai12}. While the exact tidal flux of Enceladus is still uncertain \citep{spe13}, it is certainly in excess of predictions obtained using $Q=18,000$ for Saturn \citep{mey07}.

\citetalias{lai12} also find Mimas evolving {\it inward} at a rate much faster than tidal dissipation can provide.  After careful consideration of \citetalias{lai12}, we conclude that the additional acceleration on Mimas is likely an artifact of their model. It appears that the motion of Mimas is harder to study than that of other moons, possibly because of its proximity to Saturn. However, the model is capable of correctly detecting the tidal acceleration of Tethys and Dione. 

In the present Letter, we show that the resonance of Titan and Hyperion naturally arises through tidal evolution if the tidal Q of Saturn is in the range found by \citetalias{lai12}. Our conclusions partially overlap with those of \citet{gre72} and \citet{gre73}, who also suggested tidal origin of the Titan-Hyperion resonance. As subsequent work mostly abandoned this hypothesis, we think that it needs revisiting in the context of recent results by \citetalias{lai12} and its implications for Enceladus.

\section{Analytical Estimates}

Here we use an analytical approach adapted from \citet[][pp 386-387] {md99} to estimate the amount of tidal evolution of Titan needed to resonantly excite Hyperion's eccentricity to its present value. The resonant argument for the Titan-Hyperion resonance is
\begin{equation}
\phi=4 \lambda_2 - 3 \lambda_1 -\varpi_2, 
\label{argument}
\end{equation}
where $\lambda$s are mean longitudes, $\varpi$ is the longitude of pericenter, and the subscripts 1 and 2 refer to Titan and Hyperion, respectively.
The resonant increase of eccentricity in this resonant pair is given by:
\begin{equation}
{<\dot{e_2}> \over e_2} = {n_2 a_2 \over e_2^2} {m_1 \over m_p} {F \over 3g}, 
\label{dedt}
\end{equation}
where $m$ is mass, $a$ is semimajor axis and $n$ is mean motion, with $m_p$ being Saturn's mass. The functions $g$ and $F$ are given by

\begin{equation}
g={4^2 G m_1 \over a_2^2} +  {3^2 G m_2 \over a_1^2} \ \ \ {\rm and} \ \ \ F = 4 \Bigl({dn_2 \over dt}\Bigr)_t - 3 \Bigl({dn_1 \over dt}\Bigr)_t,
\label{gandF}
\end{equation}
with the subscript $t$ referring to changes to mean motion due to tides on Saturn (rather than perturbations from the other moon), and $G$ being the gravitational constant. The mass and tidal recession of Hyperion are small compared to those of Titan and can be neglected. Therefore we get:
\begin{equation}
<\dot{e_2}>  = - {n_2 a_2 \over e_2} {m_1 \over m_p} { \dot{n_1} a_2^2 \over 16  G  m_1} = {1 \over 8 e_2} {\dot{a}_1 \over a_1},
\label{dedt2}
\end{equation}
where we used $n_1 = 4/3 \ n_2$, $\dot{n}/n = - 3/2 \ \dot{a}/a$  and $Gm_p = n^2 a^3$.
After integrating equation \ref{dedt2} (assuming a circular orbit for Hyperion before the resonance), we get:
\begin{equation}
\ln( a_1 / a_{1,0} ) = 4 e_2^2
\label{result}
\end{equation}
Since $e_2=0.1$, post-resonance-capture tidal expansion of Titan's orbit should be about 4\%, in agreement with the numerical results (Section 3). Using the expression in \citet{md99}, we connect the expansion of Titan's orbit to the $Q/k_2$ of Saturn:
\begin{equation}
{2 \over 13} \bigl( a^{13/2}_1 - a^{13/2}_{1,0} \bigr)= 3 {k_2\over Q} {m_1 \over m_p}\sqrt{G m_p} R_p^{5} \tau,
\label{tidal}
\end{equation}
where $R_p$ is Saturn's radius and $\tau$ the age of the resonance. Using Eq. \ref{result} and solving for $Q/k_2$, we get (using $P_1=2 \pi \sqrt{a_1^3 / G m_p}$):
\begin{equation}
{Q \over k_2} = {39 \pi \over 1-\exp(-26 e_2^2)} \ {m_1 \over m_p} \ \Bigl({ R_p \over a_1}\Bigr)^5 \ {\tau \over P_1}
\label{qk2}   
\end{equation}
Using $e_2=0.1$ (the approximate ``free" eccentricity of Hyperion), $\tau=4.5$~Gyr, and taking other quantities from Appendix A of \citet{md99}, we get $Q/k_2 = 3800$, in rough agreement with the results of \citetalias{lai12}. It is important to note that this calculation estimates {\it average} tidal dissipation in Saturn over its lifetime, while the \citetalias{lai12} result is derived from observations spanning about a century. Also, it is possible that the tidal $Q$ experienced by different moons can vary, due to frequency dependence \citep{gre73, fer05}. 

\section{Numerical Integration}

In order to study tidal evolution of Saturn's moons, we wrote a new symplectic integrator, SIMPL (Symplectic Integrator for Moons and PLanets). This is a combination of two Levison-Duncan type integrators \citep{lev94}, one for the planets and one for the moons of one planet. The algorithm was adapted from \citet{cha02}, and forces on the moons include mutual, solar and planetary perturbations, the planet's oblateness, planetary and satellite tides, and parametrized planetary migration. 

In the simulations shown here, the Solar System includes only Jupiter, Saturn and the Sun, we merged the moons of Saturn interior to Titan into the planet's gravity moment $J_2$, and Iapetus was omitted in some simulations but included in the others (next section). We assumed that Titan evolved tidally due to tides raised on Saturn, and satellite tides were ignored. Titan was assumed to start about 4\% interior to its present location, with low eccentricity and inclination, with Hyperion placed just outside its 4:3 resonance, also on a low eccentricity orbit (with an inclination close to the present value of $0.6^{\circ}$). We used a timestep of 0.73 days for all bodies in the system.

\begin{figure}[h]
\includegraphics[scale=1]{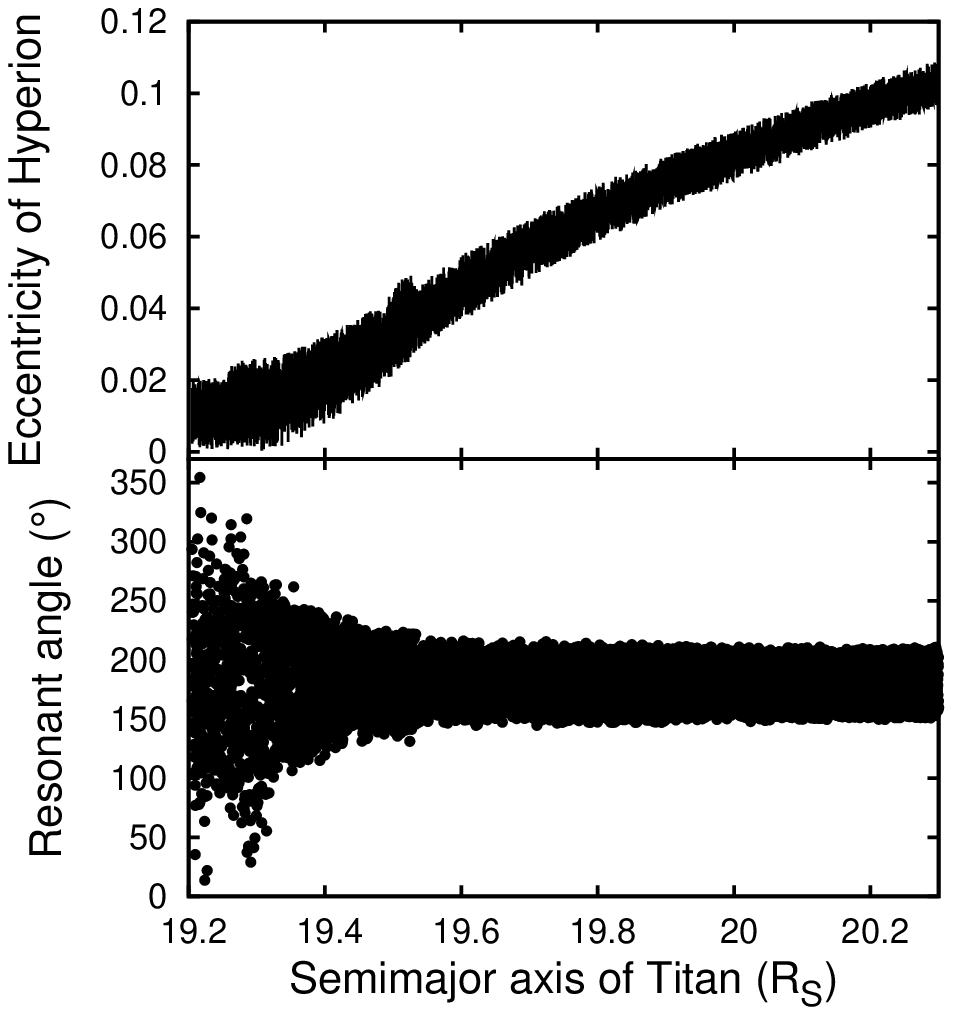}
\caption{Numerical simulation of tidal evolution into the Titan-Hyperion 4:3 resonance, accelerated about 40 times. The top panel shows the evolution of Hyperion's eccentricity as Titan migrates out, and the bottom panel shows the resonant argument $4 \lambda_H - 3 \lambda_T - \varpi_H$, where $\lambda$ and $\varpi$ stand for mean longitude and the longitude of pericenter, respectively, and the subscripts H and T refer to Hyperion and Titan.\label{hyp}}
\vspace{-.4in}
\end{figure}

Using SIMPL  we confirm that tidal evolution can naturally generate the present 4:3 Titan-Hyperion resonance. In Fig. \ref{hyp}, we show the evolution of the eccentricity of Hyperion and the Titan-Hyperion resonant argument as Titan evolves tidally from about 96\% of its semimajor axis to its present distance. The tidal $Q$ of Saturn was set to 100 in this simulation in order to execute the experiment in a manageable amount of simulation time ($<200$~Myr). This tidal evolution rate is still very slow compared to secular periods of the moons (which are measured in centuries), so we are confident that we are capturing the correct dynamics of the system.\footnote{Simulations using $Q=2$ and $Q=20$ for Saturn produce the same results as the one with $Q=100$, confirming that resonant capture in the ``slow" regime is independent of the tidal evolution rate.} The final proper eccentricity of Hyperion is about 0.1, and the resonant argument librates around $180^{\circ}$ with an amplitude of about 30$^{\circ}$. While the present libration amplitude is 50$^{\circ}$, it appears to be related to the eccentricity of Titan, which may have been excited more recently (see next section).

\section{Resonances with Other Moons}

If the tidal Q found by \citetalias{lai12} is correct, there are two new mechanisms of perturbing Titan. First, Titan itself would migrate through tides, and encounter the 5:1 resonance with Iapetus. Second, the inner moons may have gone through many cycles of migration, disruption and re-accretion, offering a wide range of possible resonances with Titan. 

The 5:1 resonance between Titan and Iapetus would have been encountered when Titan was about $0.4\%$ closer in than it is now, which is about 500 Myr ago if Saturn has $Q/k_2=$4000. We have modeled this resonance crossing and found that the orbits of both bodies are chaotic during the crossing of the resonant region, which consists of numerous sub-resonances of the 5:1 resonance. If Titan was as eccentric then as it is now, we find that the most likely outcome is eccentricity growth for Iapetus, followed by orbit crossing. If the eccentricity of Titan was low (about $0.005$), Iapetus typically survives the resonance, acquiring an eccentricity of a few percent, consistent with its present orbit ($e=0.03$). Therefore, if the Titan-Hyperion 5:1 resonance was crossed in the past (which is unavoidable for $Q/k_2 = 4000$), the current large eccentricity of Titan must postdate this resonance crossing. 

The inclination of Iapetus is affected only weakly by the 5:1 resonance crossing, with the typical change being only a degree or so (the current free inclination of Iapetus is $8^{\circ}$). Therefore, this resonance cannot constrain the timing or the source of Iapetus's inclination excitation. Interestingly, changes to the inclination of Titan can be comparable to the free inclination itself (which is $0.3^{\circ}$), making it possible that Titan's inclination was significantly modified by this resonance. 

\begin{figure}[h]
\includegraphics[scale=1]{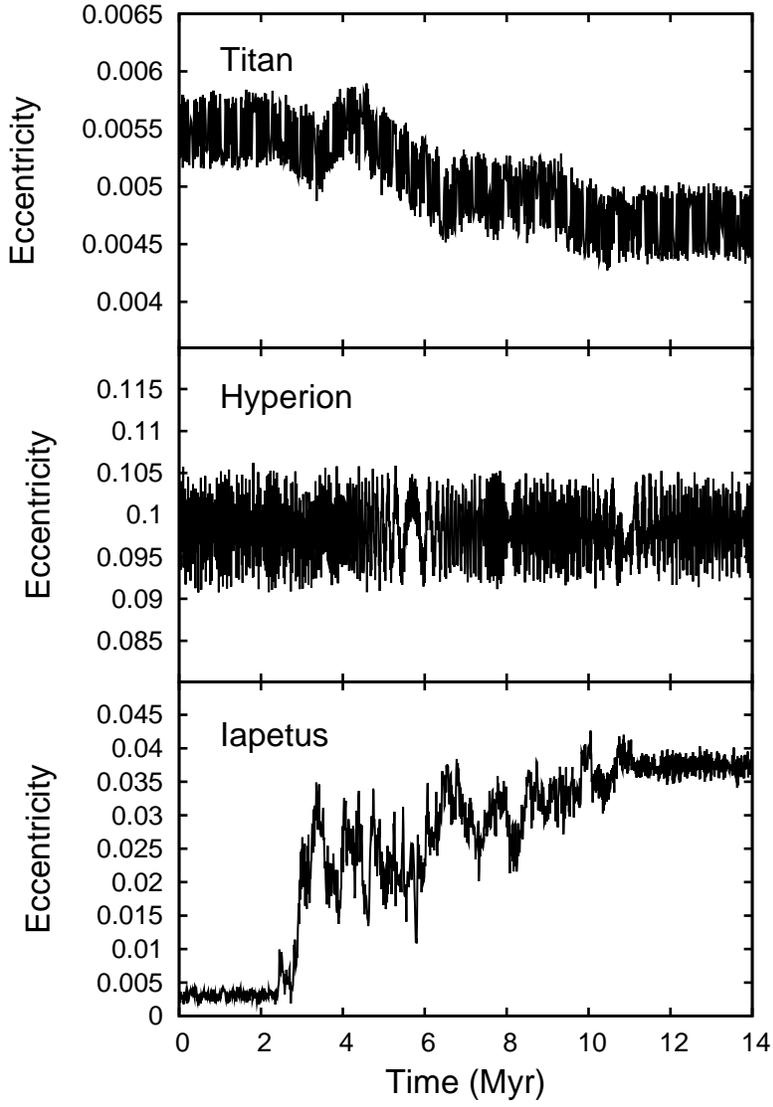}
\caption{Numerical simulation of tidal evolution through the Titan-Iapetus 5:1 resonance, using our nominal $Q/k_2=4000$. The top, middle and bottom panels plot the eccentricity of Titan, Hyperion and Iapetus.\label{hiap}}
\vspace{-.4in}
\end{figure}

Here we will concentrate on the consequences of the Titan-Iapetus resonance for Hyperion. Hyperion is much less massive than either Titan or Iapetus, so it can in principle be destabilized while the two larger moons are crossing their resonance. Destabilization of Hyperion in this resonance which {\it must} happen if the tidal Q of Saturn is low would directly falsify our hypothesis. Fig. \ref{hiap} shows the eccentricities of Titan, Hyperion and Iapetus during a simulation of the Titan-Iapetus 5:1 resonance crossing. The eccentricity of Titan was assumed to have been low ($e=0.005$) at this epoch, as high eccentricities lead to instability of Iapetus. Other orbital elements were taken to be the same as now, except that Iapetus was moved to a semimajor axis just outside the 5:1 resonance with Titan. The tidal Q of Saturn was taken to be $Q=4000$, and satellite tides were ignored.

Figure \ref{hiap} shows that there is no major effect on the eccentricity of Hyperion from the resonance. While Iapetus is affected significantly, the only consequence for Hyperion is a very small change in its forced eccentricity, which is proportional to the eccentricity of Titan. This variation of the forced eccentricity of Hyperion shows as a small decline in the amplitude of eccentricity oscillations. We conclude that the present resonant orbit of Hyperion is fully compatible with the crossing of the Titan-Iapetus 5:1 resonance in the last Gyr.

Resonances with the inner moons lead to very different results. Rhea is currently the only moon exterior to significant resonances with Titan and Hyperion (the 5:1 and 4:1 Rhea-Titan resonances). If Rhea underwent extensive tidal evolution, crossing of these resonances would significantly excite or even destabilize Hyperion (the 4:1 Rhea-Titan resonance commonly ejects Hyperion). Additionally, Rhea acquires inclination while crossing these resonances that does not get damped and is significantly in excess of its observed orbital tilt ($0.33^{\circ}$). This implies that Rhea could not have migrated continuously over billions of years \citep[cf.][]{cha11}, but that Rhea, like the other inner satellites, must have been re-accreted close to the present location much more recently. 

The mechanism of the recent disruption of the previous generation of icy moons, and (probably related) excitation of Titan is outside of the scope of this Letter. We suspect that both were caused by a semi-secular resonance between an inner moon, the Sun and Titan. This resonance is located at the distance where the apsidal precession period is one half of Saturn's orbital period, at about 7 Saturn radii (outside the present orbit of Dione). We will describe the details of this resonance in a separate publication.

\section{Conclusions}

We show that the low tidal $Q$ of Saturn found by \citet{lai12} is strongly supported by the existence of the Titan-Hyperion 4:3 resonance \citep[cf.][]{gre72}. This resonance was likely established in the early Solar System and kept evolving due to Titan's tidal recession. The 5:1 Titan-Iapetus resonance, which likely happened in the last Gyr, does not destabilize Hyperion. Rhea could not have had unbroken tidal evolution from an orbit close to the rings \citep{cha11}, as it would destabilize Hyperion when crossing the 4:1 resonance with Titan. Therefore, as first noted by \citet{gre73}, the inner moons are likely much younger than Titan, Hyperion and Iapetus. 

\acknowledgments

M. {\'C} is supported by NASA's Outer Planets Research Program award NNX11AM48G.  


\nopagebreak

\newpage

\end{document}